\documentclass[8pt]{article}  \usepackage{times}
\usepackage{graphicx}

\usepackage{color} 

\begin{document}







\twocolumn[{\Huge \textbf{The thermodynamics of general 
anesthesia\\*[0.1cm]}}\\

{\large Thomas Heimburg,$^{\ast}$ and Andrew D. Jackson\\*[0.1cm]
{\normalsize The Niels Bohr Institute, University of Copenhagen,
Blegdamsvej 17, 2100 Copenhagen \O, Denmark\\}
{\normalsize $^{\ast}$corresponding author, theimbu@nbi.dk\\*[0.3cm]}
{\normalsize keywords: anesthetics, freezing point depression, E.coli, lipid 
membranes, pressure reversal, nerves\\*[0.3cm]}

    {\normalsize It is known that the action of general anesthetics is proportional
    to their partition coefficient in lipid membranes (Meyer-Overton
    rule).  This solubility is, however, directly related to the
    depression of the temperature of the melting transition found
    close to body temperature in biomembranes.  We propose a
    thermodynamic extension of the Meyer-Overton rule which is based
    on free energy changes in the system and thus automatically
    incorporates the effects of melting point depression.  This model
    provides a quantitative explanation of the pressure reversal of
    anesthesia.  Further, it explains why inflammation and the
    addition of divalent cations reduce the effectiveness of
    anesthesia.\\*[0.5cm] }}]
\section*{Introduction} More than 100 years ago Hans Meyer in Marburg
\cite{Meyer1899} and Char\-les Ernest Overton in Z\"{u}rich
\cite{Overton1901} independently found that the action of general
anesthetics is related to their partition coefficient between water
and olive oil.  Overton performed experiments on tadpoles and recorded
the critical drug concentration at which they stopped swimming.
Assuming that the solubility of these anesthetics in olive oil is
proportional to that in biomembranes, he suggested that this critical
concentration corresponded to a fixed concentration in biomembranes.
Small molecules, as different as nitrous oxide, chloroform, octanol,
diethylether, procaine, and even the noble gas xenon, all act as
anesthetics.  Overton noted that this action is completely unspecific,
i.e. dependent only on the solubility of the anesthetic in oil and
independent of its chemical nature.  Surprisingly, this finding is
still valid for general and local anesthetics
\cite{Overton1901,Langerman1994,Kreienbuhl2002,Urban2002} but remains
unexplained.  Overton concluded that this non-specificity requires a
single mechanism based on physical chemistry and not on the molecular
structure of the drugs.  Although the close relation between
anesthetic effect and solubility in lipids led many scientists to
believe that anesthetic action is lipid-related, no model was proposed
by Meyer and Overton or by later research.  It is known, however, that
lipid melting transitions are lowered in the presence of anesthetics
\cite{Kharakoz2001}.

In the absence of a satisfactory physiological membrane mechanism,
many others prefer to view the action of anesthetics as due to
specific effects on proteins, e.g. sodium channels or
luciferase \cite{Peoples1996,Franks1994,Franks1998}.  Since
anesthetics act on nerves and the Hodgkin-Huxley theory for the action
potential is based on the opening and closing of ion channels, it
seems natural to attribute the action of anesthetics to interactions
with these channels.  Some anesthetics show a
stereospecificity indicating that the effective anesthetic
concentration (ED$_{50}$) is different for the two chiral forms even
though the partition coefficient is not affected to the same degree
\cite{Firestone1987}. In this regard, however, we note that lipid
molecules are also chiral.  While it widely believed that local
anesthetics are sodium channel blockers, a satisfactory
general model of how anesthetics act on proteins is again
lacking.  The action of anesthetics is still mysterious.  Some
lipid and protein theories on anesthesia are reviewed in
\cite{Roth1979,Peoples1996}.

The general absence of specificity and the strong correlation between
solubility in lipid membranes and anesthetic action seems to speak
against specific binding and a protein mechanism.  On the other hand,
there is clear evidence that the action of some proteins is influenced
by anesthetics.  Data on the influence of anesthetics on luciferase
and on Na- and K-channels are summarized in \cite{Firestone1986} and
suggest that the action of lipids and that of proteins are coupled in
some simple manner.  Cantor has thus proposed that all
membrane-soluble substances alter the lateral pressure in the
hydrocarbon region and thereby influence the structure of proteins
\cite{Cantor1997a,Cantor1997b,Cantor2001}.  Lee proposed a
coupling of protein function to the transition temperature of a lipid
annulus at the protein interface \cite{Lee1977b}. While such
mechanisms may provide a control of protein function, it is
nevertheless remarkable that all animals are affected to the same
degree by anesthetics, suggesting that anesthetic action is largely
independent of the specific protein composition of membranes.  (See
\cite{Overton1901}, foreword to the English edition.)  In addition to
their effect on nerves, anesthetics also change membrane properties
such as permeability and/or the hemolysis of erythrocytes
\cite{Firestone1986,Urban2002}.  This indicates the need for a more
general view of anesthetic action.

In this paper we focus on a thermodynamic description of general
anesthesia based on lipid properties.  We recognize that this can seem
heretical given the dominance of the ion channel picture.
Nevertheless, there are a variety of reasons for considering a
macroscopic thermodynamic view.  The striking fact that noble gases
can act as general anesthetics speaks against specific binding to
macromolecules.  In particular, the Meyer-Overton rule would require
all anesthetics to have exactly the same partition coefficient between
lipid membrane and protein binding sites for all relevant proteins.
It is difficult to imagine that nature provides binding sites for such
a variety of molecules on the same protein in precisely such a manner
that binding affinity is independent of chemical nature
\cite{Miller2006}.  An acceptable description should account for this
evident lack of specificity, and this suggests the utility of
thermodynamic arguments.  Moreover, it is to be emphasized that
thermodynamics is not inimical to microscopic (e.g., ion-channel)
descriptions of the same phenomena.  No one would claim, for example,
that the manifest successes of thermodynamics in describing the
properties of real gases in any way contradicts the fact that they are
composed of interacting atoms.  Thermodynamics rather recognizes that
many macroscopic phenomena are independent of such microscopic details
and that a large number of microscopic systems can display features
which are both qualitatively and quantitatively susceptible to more
generic methods.  Precisely the absence of detail means that
thermodynamic approaches are often capable of making testable
quantitative predictions which are often inaccessible to or obscured
by more microscopic models.  Thus, we wish to propose a simple
thermodynamic explanation of the Meyer-Overton rule based on the
well-known physical chemical phenomenon of freezing-point depression.
We will show that this picture has the benefit of providing an
immediate and intuitive mechanism for the pressure reversal of
anesthesia as a consequence of the pressure-induced elevation of the
melting point in lipid membranes and can explain the effects of
inflammation and divalent cations on anesthetic action.

\section*{Materials and Methods}
Lipids were purchased from Avanti Polar Lipids (Birmingham, AL) and
used without further purification. Octanol was purchased from Fluka 
(Buchs, Switzerland). Multilamellar lipid dispersions
(5mM, buffer: 2mM Hepes, pH 7.4, octanol concentration adjusted) were
prepared by vortexing the lipid dispersions above the phase transition
temperature of the lipid. We also performed experiments with 
halothane and other anesthetics that yielded results similar to those 
of octanol. These data are not shown here.\\

\textit{E.\,coli} bacteria (XL1 blue with tetracycline resistance) and
bacillus subtilis were grown in a LB-medium at 37\,$^{\circ}$C. The
bacterial membranes were then disrupted in a French Press at 1200 bar
(Gaulin, APV Homogeniser GmbH, L\"{u}beck, Germany) and centrifuged at
low speed in a desk centrifuge to remove solid impurities.  The
remaining supernatant was centrifuged at high speed in an Beckman
ultracentrifuge (50000 rpm) in a Ti70 rotor to separate the membranes
from soluble proteins and nucleic acids.  This membrane fraction was
measured in a calorimeter.  Lipid melting peaks and protein unfolding
can easily be distinguished in pressure calorimetry due to their
characteristic pressure dependences.  The pressure dependence of lipid
transitions is much higher than that of proteins and nearly
independent of the lipid or lipid mixture \cite{Ebel2001}.  Further,
in contrast to lipid transitions, the heat unfolding of the proteins
is not reversible.  More details regarding the \textit{E.\,coli}
measurements are given in an MSc.\ thesis \cite{Pollakowski2004} and
will be published elsewhere.  

Heat capacity profiles were obtained using a VP-scanning calorimeter
(MicroCal, Northampton, MA) at scan rates of 5 deg/hr (lipid
vesicles) and 30\,deg/hr for \textit{E.\,coli} membranes.\\

To calculate the theoretical heat capacity profiles we used ideal
solution theory, described in \cite{Lee1977}.  It was assumed that the
anesthetic is ideally miscible with the fluid phase and immiscible in
the gel phase.  These assumptions are in agreement with experiment.
Due to the partition coefficient in the membrane most of the
anesthetic is found in the membrane (P=200 for DPPC membranes
\cite{Jain1978}) if the amount of the aqueous phase is small.  Under
such conditions, the anesthetic concentration in the fluid phase
changes when lowering the temperature below the onset of the melting
transition.  The chemical potentials of the gel and the fluid lipid
membrane are given by
\begin{equation}
    \mu^g=\mu_{0}^g\qquad,\qquad\mu^f=\mu_{0}^f+RT\ln(1-x_{A})
    \label{0.1}
\end{equation}
where $x_{A}$ is the molar fraction of anesthetics in the membrane.
$\mu_{0}^g$ and $\mu_{0}^f$ are the standard state chemical potentials
that obey the relation
\begin{equation}
    \mu_{0}^f-\mu_{0}^g=\Delta H\left(1-\frac{T}{T_{m}}\right)
    \label{0.2}
\end{equation}
with the excess enthalpy of the transition, $\Delta H$, and the melting
temperature, $T_{m}$.  With these assumptions, one can calculate phase
boundaries and melting point depression (see next section).  Using the
lever rule one can deduce the the relative fractions of gel and fluid
phase as a function of temperature \cite{Lee1977}.  When the fraction
of fluid phase, $f_{fluid}$, is multiplied with the excess melting
enthalpy, $\Delta H$, one obtains the enthalpy as a function of
temperature, $\Delta H(T)=f_{fluid}\cdot\Delta H$.  The excess heat
capacity is the derivative of this function.  For details see also
\cite{Heimburg2007a}.\\

\section*{Theory and results}
\paragraph*{The unspecific effect of anesthetics and other small
solutes on lipid melting transitions} Biological membranes are 
\linebreak[4] known
to undergo a phase transition from a low-temperature solid-ordered (SO
or gel) phase to a liquid-disordered (LD or fluid) phase at
temperatures slightly below physiological temperature.  This
transition involves a volume change of $\approx$\,$4$\% and an area
change of $\approx$\,$25$\%.  It is also known empirically that nerve
pulses are accompanied by density and heat \cite{Ritchie1985}
changes consistent with forcing the lipid mixture through
$\approx$\,$85$\% of this phase transition
\cite{Kaminoh1991,Kaminoh1992}.  When supplemented by the empirical
observation that the sound velocity in lipid mixtures increases with
frequency, this fact leads to the robust prediction that localized
piezo-electric pulses (or ``solitons'') can propagate stably in
biological membranes \cite{Heimburg2005c,Lautrup2006}.  The lipid
melting transition is essential for the existence of solitons.  In the
transition from the LD to the SO phase, membranes become more
compressible and also permeable for ions and molecules
\cite{Papahadjopoulos1973,Boheim1980,Cruzeiro1988}.  The biological
membrane thus resembles a spring that becomes softer upon compression.
This non-linearity is necessary for the formation of solitons, which
can propagate in cylindrical membranes without distortion even in the
presence of significant noise.  Such a description can account
naturally for the reversible heat and mechanical features of nerve
pulses and also predicts a pulse propagation velocity of
$\approx$\,$100$\,m/s, which is comparable to that in myelinated
nerves.

Given the existence of a lipid phase transition and its possible
biological relevance, it is tempting to speculate that it plays a
functional role in unspecific anesthetic effects and that it is
central to understanding the Meyer-Overton rule.  The basis for such
speculation is elementary.  The introduction of any solute (i.e.,
anesthetic) into membranes leads to a lowering of the temperature of
the melting transition which is proportional to the molar
concentration of the solute and largely independent of its chemical
nature.

\begin{figure}[htb!]
    \begin{center}
	\includegraphics[width=8cm]{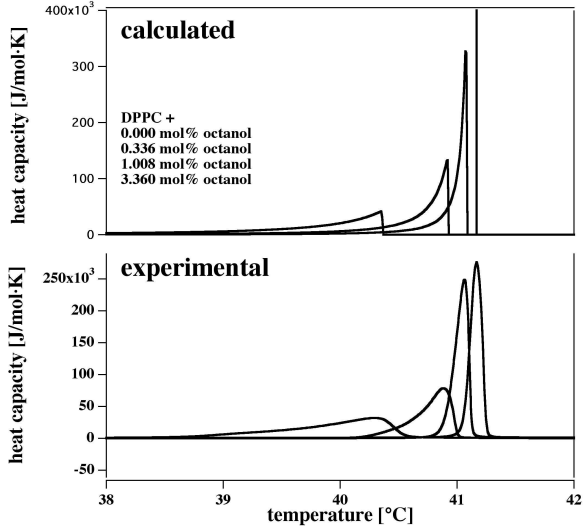}
	\parbox[c]{8cm}{ \caption{\textit{The effect of octanol on
	the phase transition of DPPC vesicles.  Bottom: Calorimetric
	data with various octanol concentrations in the membrane.
	Top: Calorimetric profiles calculated for the same membrane
	concentrations of a solute assuming ideal mixing in the fluid
	phase and no mixing in the gel phase.  The high temperature
	end of the transition profile corresponds to the temperature
	calculated for melting point depression.  The calculation
	assumes a finite bulk fluid phase.  This leads to an
	accumulation of anesthetics in the fluid phase as the
	temperature is lowered and to an asymmetric broadening of the
	c$_{p}$ profile.}
	\label{fig:Figure1}}}
    \end{center}
\end{figure}

Small molecules, peptides and proteins are not in general readily
soluble in the SO-phase due to its crystalline structure.  They are
much more soluble in the LD phase.  This leads to a reduction of
melting points, demonstrated in Fig.\,\ref{fig:Figure1} for the
artificial lipid DPPC in the presence of the local anesthetic octanol.
This effect is known as freezing point depression \cite{Silbey2001}.
For example, the solubility of NaCl is high in water and low in ice.
Thus, salt lowers the freezing point of water.  This effect is due to
the difference in mixing entropy of the ions in water and ice.  For
low solute concentrations and with the reasonable assumptions of
perfect miscibility of an anesthetic in the LD phase and immiscibility
in the SO phase, one arrives at the following relation between melting
point depression and solute concentration
\cite{Silbey2001,Kaminoh1992}:
\begin{equation}
    \Delta T_{m}=-\left(\frac{RT_{m}^2}{\Delta H}\right)\,x_{A} \ ,
    \label{eq:1}
\end{equation}
where $x_{A}$ is the molar fraction of anesthetic in the membrane,
$\Delta H$ is the lipid melting enthalpy (approximately $35$\,kJ/mol
for DPPC) and $T_{m}$ is the lipid melting temperature ($314.3$\,K for
DPPC,and $295$\,K for native \textit{E.coli} membranes).  An
anesthetic concentration of 1 mol\% in the fluid membrane leads to
$\Delta T_{m}=-0.24$\,K.\\

The heat capacity can be calculated as a function of temperature for
various solute concentrations using ideal solution theory
\cite{Lee1977} with the assumption of complete insolubility in the
solid phase (Fig.\,\ref{fig:Figure1}, top).  The peak in this figure
corresponds to the phase transition.  We have assumed a small amount
of the water phase (as used experimentally) and an accumulation of
anesthetics in the fluid phase.  This leads to the broadening of the
profiles, which are remarkably similarity to experimental results
obtained for DPPC vesicles in the presence of various as shown in the
lower panel.  The quality of this agreement indicates
that thermodynamic properties (e.g., the Gibbs free energy) of the
lipid mixture are dominated by the lipid phase transition.  We will
make use of this fact below.\\

The anesthetic concentration in membranes at critical do\-sage can be
calculated using the partition coefficient, $P$, extracted from data
collected in \cite{Firestone1986} for water-soluble anesthetics and
tadpole anesthesia.  Solvents include octanol/water, PC or
egg-PC/water, and erythrocyte or PC+cholesterol/water.  This data
includes $28$ separate solute/solvent combinations for which the
partition coefficients vary by a factor of $7000$.  Log-log plots of
$P$ versus ED$_{50}$, defined as the concentration in molar units at
which 50\% of tadpoles are immobilized, reveal that the data is
consistent within error with a straight line of slope $-1$.  The
Meyer-Overton rule is fulfilled independent of the reference system.
Additional modern confirmation of the Meyer-Overton rule can be found
in \cite{Kharakoz2001} and \cite{Overton1901} (foreword to the English
edition).  The partition coefficient of membranes high in cholesterol
is smaller than that of cholesterol-free membranes.  Since nerves have
a relatively low cholesterol content (i.e., less than 10\%), we will
use the partition coefficient in PC or Egg-PC as a reference in the
following.  The assumption of linear dependence of anesthesia on the
partition coefficient is an idealization.  Characteristic deviations
are roughly a factor of two (comparable to that found for different
chiral forms) and are not large given the full range of partition
coefficients spanned by the data.  We use only data for tadpole
narcosis, where the signature of anesthesia is unambiguous.  A
least-squares fit yields
\begin{equation}
   \ln(P) = -3.38 - \ln({\rm ED}_{50})\qquad\mbox{for PC or
   egg-PC/water.}
   \label{eq:1b}
\end{equation}
The molar fraction of anesthetics in the fluid membrane at anesthetic
dose is readily determined using eq.\,\ref{eq:1b} as
\begin{equation}
    x_{A}=P ({\rm ED}_{50}) V_{l} \ ,
    \label{eq:1c}
\end{equation}
where the molar volume of fluid lipids, $V_{l}$, is taken here to be
0.750 l/mol.  This yields a membrane concentration of about 2.6 mol\%
of anesthetics in egg-PC membranes independent of anesthetic.
According to eq.\,\ref{eq:1} this corresponds to $\Delta T_{m}\approx
-0.60$\,K at anesthetic dose for tadpoles.  Kharakoz
\cite{Kharakoz2001} obtained $\Delta T_{m}\approx -0.53$\,K directly
from data for a series of alkanols, which corresponds to an anesthetic
concentration of 2.3 mol\% in membranes.  The striking agreement of
these results indicates that the freezing point depression of
eq.\,\ref{eq:1} provides an adequate description of the experimental
shifts in T$_{m}$.

The phenomenon of melting point depression, illustrated here for
octanol, allows us to re-express the Meyer-Overton rule as: The
efficacy of a general anesthetic is determined by its ability to lower
the lipid transition temperature.  Deviations from the rule usually
indicate that the assumptions of ideal mixing in the fluid phase
and/or no mixing in the solid lipid phase are not quantitatively
correct.  In particular, small noble gases atoms are also likely to
dissolve in the solid lipid phase.  Large anesthetics may display
phase behavior on their own, i.e., they may not mix ideally in fluid
lipids.  In the following, we will be concerned with anesthetics that
do follow the Meyer-Overton rule.  It is our expectation that the
thermodynamic consequences of Cantor's model
\cite{Cantor1997a,Cantor1997b,Cantor1999b} will be consistent with our
picture.

\begin{figure}[htb!]
    \begin{center}
	\includegraphics[width=8cm]{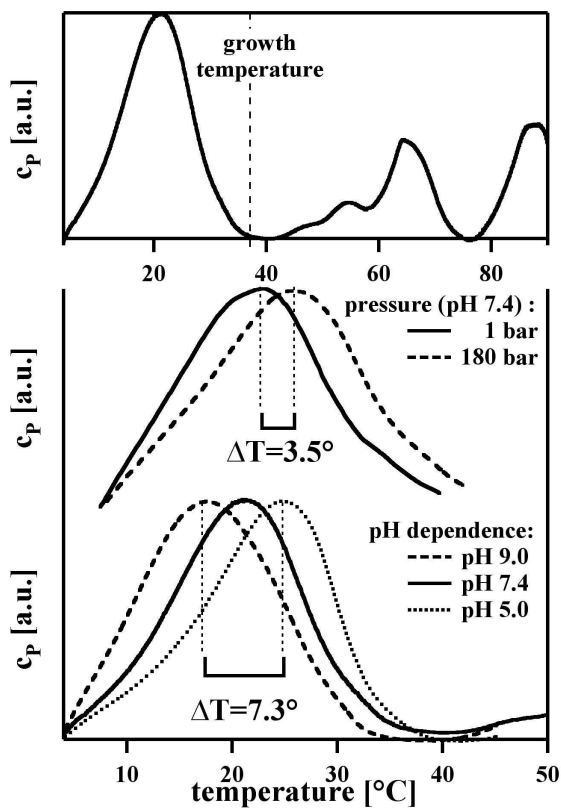}
	\parbox[c]{8cm}{ \caption{\textit{Heat capacity profiles of
	native E.coli membranes.  Top: At $37^{\circ}$\,C. The growth
	temperature is indicated.  The large peak below growth
	temperature corresponds to lipid melting.  The smaller peaks
	above growth temperature correspond to protein unfolding.
	Center: The heat capacity as a function of $T$ at a
	hydrostatic pressure of $1$ bar and $180$ bar.  The pressure
	induced shift is $\approx$$3.5$\,K. Bottom: Heat capacities
	for the same membranes at various pH values.  The transition
	temperature increases by about $5.3$\,K when the pH is reduced
	from $7.4$ to $5.0$.  Scans were halted at
	$40$-$43\,^{\circ}$\,C to prevent protein unfolding.}
	\label{fig:Figure3}}}
    \end{center}
\end{figure}

\paragraph*{The effect of pressure on transitions} Anesthetics action
can be reversed by hydrostatic pressure \cite{Halsey1975}.  In
tadpoles, a bulk pressure of 140-350 bars reverses the action of
3\,-6\, vol\% etha\-nol narcosis \cite{Johnson1950}.  It has
been suggested that this effect is related to the chain melting
transition of lipid membranes \cite{Trudell1975,Lee1977,Kamaya1979,Galla1980}. 
Melting transitions
move to higher temperatures with bulk pressure, $\Delta p$, due to the
fact that the volume of membranes in the SO phase is reduced by about
4\%.  The shift is given as
\begin{equation}
    \Delta T_{m}=\gamma_{v}\Delta p\,T_{m} \ .
    \label{eq:2}
\end{equation}
Here, $\gamma_{v}=7.8\cdot 10^{-10}$\,m$^2$/N is constant within
errors for a variety of artificial and biological membranes
\cite{Heimburg1998,Ebel2001}.  See also Fig.\,\ref{fig:Figure3}.
Luciferase, which is regarded as a model protein for general anesthesia,
does not display pressure reversal \cite{Moss1991}.

\paragraph*{Free energy changes} Although the internal energy (or
enthalpy) of a lipid membrane above the melting temperature is
insensitive to changes in the lipid transition temperature, the
associated free energy change has significant temperature dependence.
Given the lipid melting enthalpy, $\Delta H$, the entropy change
associated with the transition is $\Delta S = \Delta H/T_m$.  The
difference between the free energies of the LD and SO phases at a body
temperature $T > T_m$, $\Delta G(T)$, is thus given as
\begin{equation}
    \Delta G( T ) \approx \Delta H \left(\frac{T_{m}-T}{T_{m}}\right)
    \ ,
    \label{eq:1d}
\end{equation}
which is explicitly sensitive to changes in $T$.  Including the
effects of anesthetics and a hydrostatic pressure, this difference in
the Gibbs free energy for membranes becomes
\begin{equation}
    \Delta G(T , \Delta p ) \approx \Delta H
    \left(\frac{T_m-T}{T_{m}}-\frac{RT}{\Delta
    H}x_{A}+\gamma_{V}\Delta p \frac{T}{T_m}\right) \ ,
    \label{eq:1e}
\end{equation}
where $T_{m}$ is the melting temperature of the membrane in the
absence of anesthetics and $\Delta p$ is the excess hydrostatic
pressure.  Obviously, eq.  (\ref{eq:1e}) can be extended to include
the effects of other relevant intensive thermodynamic variables such
as the chemical potentials of hydrogen ions or calcium.  The
Meyer-Overton rule indicates that the free energy difference is
increased by $\approx$\,5\% by the addition of a critical dose of
anesthetics.  Since this energy must be supplied from chemical
sources, it is natural to postulate that equal values of $\Delta G ( T
, \Delta p)$ will produce equal anesthetic effect.  This postulate
represents an extension of the Meyer-Overton rule, and
eq.\,\ref{eq:1e} leads to a variety of specific and quantitative
predictions regarding anesthetic action and other phenomenon governed
by this phase transition.

\paragraph*{Pressure reversal of anesthesia} From eq.\,(\ref{eq:1e}),
the pressure required to reverse the action of an anesthetic is
\begin{equation}
   \Delta p \approx \frac{1}{\gamma_{v}}\frac{RT_{m}}{\Delta H} \,
   x_{A} \ .
   \label{eq:3}
\end{equation}
The hydrostatic pressure required to reverse the action of anesthetics
on the phase transition is $9.6$\,bar/mol\% using the values of
$\Delta H$ and $T_{m}$ appropriate for DPPC. Here and below, the
reversal of anesthetic effect would be complete if the heat capacities
shown in figure \ref{fig:Figure1} were of zero width.  In practice,
this reversal is only approximate due to the broadening of the
profiles.
Pressure reversal of anesthesia was first demonstrated by Johnson and
Flagler \cite{Johnson1950}.  They anesthetized tadpoles in 3-6 vol\%
ethanol.  A hydrostatic pressure of 140-350 bars was found to reverse
anesthesia.  According to \cite{Firestone1986}, 190mM of ethanol (1.1
vol\%) in the aqueous phase is necessary for tadpole narcosis.  This
means that about 3-6 times the anesthetic ethanol concentration was
used in \cite{Johnson1950}.  The concentration of ethanol in the
membrane in Johnson and Flagler's experiments was therefore 7.5-15
mol\%.  According to eq.\ref{eq:1}, these concentrations correspond to
lowering $T_m$ by 1.8-3.6 K. From eq.\ref{eq:3}, the pressure
necessary to reverse this anesthetic effect is 72-148 bars.
Considering the uncertainty of the partition coefficient for real
biological membranes (which depends on the precise lipid mixture),
this is remarkably close to the order of the values found by
\cite{Johnson1950}.  The fact the pressure increases $T_{m}$ may be
related to the observation that nerves fire spontaneously at high
pressures \cite{Kendig1978}.

\paragraph*{Effects of pH and salts} Ions also change the free energy.
Some 10\% of the lipids of biological membranes are negatively
charged, primarily on the inner membrane.  At lower pH, some of these
charges are protonated, and the electrostatic potential of the lipid
membrane is reduced.  Complete protonation increases the melting
temperature by approximately $20$\,K. The effects of pH and ionic
strength on melting transitions have been carefully investigated by
\cite{Traeuble1974,Traeuble1976}.  While these effects depend on the
precise composition of the membrane and on ionic strength, they can be
calculated using Debye-H{\"u}ckel theory or determined empirically.
For example, the temperature of the melting transition in native
\textit{E.coli} membranes (in the pH range between 5 and 9) is raised
by about $1.8$ degrees if pH is lowered by one unit
(Fig.\ref{fig:Figure3}).  This shift is approximately that which is
produced by $72$ bars hydrostatic pressure.  Interestingly, it is
known that inflammation leads to the failure of anesthesia.  The
related lowering of pH in inflamed tissue, i.e., on the order of 0.5
pH units \cite{Punnia-Moorthy1987}, is widely assumed to be
responsible.  According to the above, the lowering of pH from $7$ to
$6.5$ leads to $\Delta T_{m}=+0.9$\,K, which is sufficient to reverse
the action of anesthetics at the typical critical dose corresponding
to $\Delta T_{m}=-0.6$\,K.

Salts can also effect the melting transition through, e.g., the
binding of divalent cations such as Mg$^{++}$ and Ca$^{++}$.  These
ions shift the melting temperatures of both charged and uncharged
lipids to higher temperatures.  The presence of such ions thus lowers
the effectiveness of anesthetics, and appropriate functions of pH and
salt concentration should be added to the right side of
eq.(\ref{eq:1e}).

\paragraph*{Temperature effects} Many processes in biology respond 
directly to temperature changes.  Therefore it can be difficult to
isolate individual temperature effects in vivo.  It has been shown by
\cite{Spyropoulos1961,Kobatake1971} that cooling can trigger the
action potential whereas heating inhibits the nerve pulse.  If the
body temperature is changed from $T$ to $T+\Delta T$, the Gibbs free
energy of the membrane also changes.  The action of anesthetics can be
reversed by changing body temperature by
\begin{equation}
    \Delta T\approx-\left(\frac{RT_{m}^2 }{\Delta H}\right) x_{A}\ .
    \label{eq:1f}
\end{equation}
The effect of $2.6$ mol\% anesthetics is thus reversed by a $0.6$\,K
reduction of the body temperature for the parameters of DPPC
membranes.  Interestingly, a well-known finding in clinical
anesthesiology is hypothermia leading to a lowering of the
body-temperature during narcosis \cite{Abelha2005}.  Conversely, the
same arguments say an equal rise in body temperature, e.g., by fever,
should produce the same effects as a critical anesthetic does.  Since
this is not the case, a rise in body temperature must be accompanied
by other thermodynamic changes which tend to counter this increase in
the free energy difference (e.g. pH changes) if our thermodynamic
picture is to be maintained.  Further, a lowering of the temperature
below the phase transition temperature ($\Delta T$ more than $-15$\,K)
would lead to a complete cessation of nerve activity as found in
clinical experiments \cite{Katz1997}.  Note that the chemical
composition of lipids can also change in response to changes in other
thermodynamic variables.  It is well documented that the lipid
composition and the melting temperatures of bacterial membranes change
as a response to changing growth temperature (e.g.
\cite{vandeVossenberg1999}. E.coli membranes grown at different
temperature shift their melting temperatures to maintain a constant
distance to growth temperature (unpublished data from our
laboratory)).

\section*{Conclusion} We have proposed an elementary thermodynamic
description of the action of general anesthesia according to which
constant anesthetic effects are predicted whenever external
thermodynamic variables (e.g., solute concentration, pressure,
temperature, pH and salt concentration) are adjusted to maintain
constant values of the free energy difference between the liquid and
gel phases of lipid membranes.  We have demonstrated that the heat
capacity and free energy of biomembranes are dominated by the
well-known membrane melting transition, which is thus seen to play a
central role in anesthetic action.  Indeed, the effect of an
anesthetic is due solely to its ability to depress the melting point
of lipid membranes, which depends on its solubility in lipid mixtures
but is otherwise independent of its chemical nature.  The basis for
the familiar Meyer-Overton rule thus lies in the thermodynamics of
biological membranes in general and the properties of the lipid phase
transition in particular.  The lowering of the membrane melting point
results in a change of the free energy of the lipid membrane, which is
proportional to the difference between body temperature and the
melting temperature of the membrane.  This temperature difference,
which is on the order of $15$\,K, is to be compared with the
shift in melting point temperature of $\approx$\,-0.6\,K at a
typical critical anesthetic dose.  Anesthetic effect can be reversed
in a quantitatively predictable manner by any mechanism that raises
the transition temperature and restores the free energy difference to
its original value.  Such mechanisms include hydrostatic pressure, a
decrease of pH, an increase of calcium concentration, or the lowering
the body temperature.  (The hydrostatic pressure necessary to
reverse anesthesia is on the order of $24$ bars, the pH change on the
order of $0.4$ pH units, and the hypothermic reversal of anesthesia is
about $0.6$ K.)  While these effects are well-documented,
they have not previously been placed in common framework.  Although we
do not question the importance of a better understanding of the
microscopic mechanisms underlying general anesthesia, these results
support the view that the thermodynamics of the lipid liquid-gel
transition is important for understanding the macroscopic effects of
general anesthetic action.  Finally, we note that a variety of
biological phenomena, including fusion and membrane permeability, may
reasonably be assumed to have a similar connection to this phase
transition and that such assumptions can be tested using approaches
similar to those presented here.

\paragraph*{Acknowledgment}
We thank Prof.\ Benny Lautrup (Niels Bohr Institute) for
critical discussions.  We also thank D.\ Pollakowski (Niels Bohr
Institute) and Dr.\ M.\ Konrad (G\"{o}ttingen) for permission to use
their data on \textit{E.coli} membranes prior to its detailed
publication.  We also thank Dr.  D.\ Kharakoz for making his 2001
paper available to us.

\footnotesize
\bibliographystyle{biophysj2005}

\end{document}